\documentclass[twocolumn,showpacs,preprintnumbers,amsmath,amssymb,prb]{revtex4}

\usepackage{graphicx}
\usepackage{dcolumn}
\usepackage{bm}

\begin{document}

\preprint{}

\title{ Universal behavior of the electron g-factor in 
GaAs/AlGaAs quantum wells}

\author{I. A. Yugova$^{1,2,\dag}$, A. Greilich$^1$, D. R. Yakovlev$^{1,3}$, A. A. Kiselev$^4$, M. Bayer$^1$, \\
V. V. Petrov$^2$, Yu. K. Dolgikh$^2$, D. Reuter$^5$ and A. D. Wieck$^5$}

\affiliation{$^1$ Experimentelle Physik II, Universit\"{a}t Dortmund,
44221 Dortmund, Germany}
\affiliation{$^2$Institute of Physics, St.-Petersburg State
University, St.-Petersburg, 198504, Russia}
\affiliation{$^3$ A. F. Ioffe Physico-Technical Institute, Russian Academy of Sciences, 194021 St. Petersburg, Russia }
\affiliation{$^4$ Department of Electrical and Computer Engineering, North Carolina State University, Raleigh, North Carolina 27695-7911, USA}
\affiliation{$^5$ Angewandte Festk\"{o}rperphysik, Ruhr-Universit\"{a}t 
Bochum, D-44780 Bochum, Germany}

\date{\today}

\begin{abstract}
{ The Zeeman splitting and the underlying value of the $g$-factor 
for conduction band electrons in GaAs/Al$_{x}$Ga$_{1-x}$As 
quantum wells have been measured by spin-beat spectroscopy based 
on a time-resolved Kerr rotation technique. The experimental data 
are in good agreement with theoretical predictions. The model 
accurately accounts for the large electron energies above the 
GaAs conduction band bottom, resulting from the strong quantum 
confinement. In the tracked range of optical 
transition energies $E$ from 1.52 to 2.0~eV, the electron $g$-factor 
along the growth axis follows closely the universal 
dependence 
$g_{||}(E) \approx –0.445 + 3.38(E-1.519)-2.21(E-1.519)^2$ (with 
$E$ measured in eV); and this universality also embraces 
Al$_{x}$Ga$_{1-x}$As alloys. The in-plane $g$-factor component 
deviates notably from the universal curve, with the degree of 
deviation controlled by the structural anisotropy. }
\end{abstract}


\maketitle

\section{Introduction}
The Lande or $g$-factor of carriers in a solid is one of the fundamental properties 
of this material \cite{Abragham, Awsch, Kosaka}. For conduction band electrons 
in semiconductors and semiconductor heterostructures it may deviate strongly 
from the free electron $g$-factor in vacuum $g_0=+2.0023$ due to the spin-orbit 
interaction, e. g. it is -0.44 in GaAs, -1.64 in CdTe and can be as large as -51 in 
the narrow band gap semiconductor InSb \cite{Landolt}.

Invention of spintronics has increased the interest in spin manipulation in 
semiconductor heterostructures \cite{Awsch, Zut04} and therefore in control of 
the carrier $g$-factors. GaAs/Al$_{x}$Ga$_{1-x}$As heterostructures are very 
suitable for this purpose as with increasing carrier confinement the electron 
$g$-factor's absolute value decreases and even crosses zero. Therefore, the Zeeman 
splitting can be fully suppressed by a proper choice of structure design parameters and/or 
external conditions like strain, temperature, electric field, and orientation of the 
structure in external magnetic field \cite{Salis01, Jiang01, Oestreich95}. 

The electron $g$-factor in GaAs/Al$_{x}$Ga$_{1-x}$ quantum well (QW) 
structures has been measured by various experimental techniques such as spin-flip 
Raman scattering \cite{Sapega92}, optical orientation \cite{Snelling91, 
Ivchenko92_hmfsp, Kalevich92_zhetf, Kalevich95_ftt}, optically detected 
magnetic resonance \cite{Baranov93, vanKesteren}, spin quantum beats in 
emission \cite{Heberle94, Hannak95, Amand97, LeJeune97}, in absorption 
\cite{Bar-Ad92}, or in Kerr rotation \cite{Malinowski00_sss, 
Malinowski00_prb}. However, most of these studies have been limited to certain 
widths of the quantum wells and only in a few papers the well width dependence 
of the $g$-factor has been reported \cite{Snelling91, Malinowski00_prb, 
Hannak95, LeJeune97}. 

The first concise analysis of the electron g factor in QWs was performed in the 
framework of the Kane model \cite{Ivchenko92_ftp}, followed by more detailed 
considerations \cite{Ivchenko97_ssc, Kiselev_PRB98, Kiselev_PhysStatSol99, 
Ivch04}. The model calculations predict a strong variation of the $g$-factor value, 
including a sign reversal in the GaAs/Al$_{x}$Ga$_{1-x}$ heterosystem, and of 
its anisotropy with well width. Both quantities are controlled by the strength of 
electron confinement in the QWs determined mostly by the well width and barrier 
height. These theoretical predictions were further substantiated by experimental 
data \cite{Sirenko, Kiselev_JCrystGrowth98, Malinowski00_prb, LeJeune97}. 
The published results are commonly presented as a dependence of the $g$-factor 
value on the QW width, i.e. a set of different dependencies corresponding to 
different barrier heights (controlled by the Al content) is required to cover the 
whole range of possible QW structures. In CdTe/Cd$_{1-x}$Mg$_{x}$Te 
heterostructures, whose band structure is similar to that of 
GaAs/Al$_{x}$Ga$_{1-x}$ structures, to a good approximation a {\em 
universal} dependence of the electron $g$-factor on the heterostructure band gap 
(i.e., on the energy of the optical transition between the ground states of confined 
electrons ($e1$) and holes ($hh1$)) has been reported \cite{Sir97}: the $g$-factor 
was sensitive mostly to the value of the band gap itself, but not to the way how 
this gap has been obtained by the structure's design parameters.

The goal of this paper is to check experimentally and theoretically, whether this 
universality can be extended to GaAs/Al$_{x}$Ga$_{1-x}$ heterosystems, and if 
yes, what the origins of this universality are. We present experimental results for 
the time-resolved pump-probe Kerr rotation, which allow us to determine the 
transverse component $g_{\perp}$ of the electron $g$-factor with high accuracy 
from the frequency of the spin precession in an external magnetic field. Model 
calculations accounting for the $k \cdot p$ interaction between lowest 
conduction band and the upper valence bands have been performed for the 
longitudinal and transverse components of the electron $g$-factor in structures 
with the Al content varied from 0 to 0.45. We found that for the longitudinal 
component a universal dependence is strongly supported. At the same time, the 
transverse component deviates notably from a universal curve, with the degree of 
deviation controlled by the structural anisotropy.

\section{Experimental details}

The GaAs/Al$_{x}$Ga$_{1-x}$ heterostructures have been grown by molecular-beam 
epitaxy on (100) oriented GaAs substrates. The sample parameters are collected 
in Table~\ref{tab:table1}. Samples $\# 1$ and $\# 2$ consist of 
several single QWs of different width separated by 50~nm thick 
Al$_{x}$Ga$_{1-x}$As barriers to prevent electronic coupling between the 
wells. Samples $\# 3$, $\# 4$ and $\# 5$ contain only one QW. All structures are 
nominally undoped, but due to residual doping of the barriers presence of 
background electrons in the QWs has been established from the observation of 
emission of negatively charged excitons. The background electrons density does 
not exceed $5 \times 10^9$~cm$^{-2}$.

For optical measurements the samples were immersed in pumped liquid helium at 
a temperature of 1.6~K, and magnetic fields up to 7~T were applied perpendicular 
to the structure growth axis (Voigt geometry). The structures were characterized 
by means of photoluminescence (PL) excited by 532~nm laser light with an 
excitation density below 0.3~W/cm$^2$.
\begin{table*}
\caption{\label{tab:table1} Parameters of the studied GaAs/Al$_{x}$Ga$_{1-
x}$As QW structures.}
\begin{ruledtabular}
\begin{tabular}{ccccc}
 Sample, Al content & QW width (nm) & transition & Laser energy 
(eV)&$|g_{\perp}|$\\
\hline
$\# 1$ (p343), x=0.33 & 14.3 & $e1-hh1$ & 1.530 & 0.34 $\pm$ 0.01 \\
 & 10.2 & $e1-hh1$ & 1.543 & 0.27 $\pm$ 0.01 \\
 & 7.3 & $e1-hh1$ & 1.569 & 0.13 $\pm$ 0.01 \\
 & 4.2 & $e1-hh1$ & 1.622 & 0.06 $\pm$ 0.004 \\
\hline
$\# 2$ (p340), x=0.34 & 17.2 & $e1-hh1$ & 1.527 & 0.40 $\pm$ 0.01 \\
 & 13 & $e1-hh1$ & 1.535 & 0.33 $\pm$ 0.01 \\
 & 13 & $e1-lh1$ & 1.542 & 0.33 $\pm$ 0.01 \\
 & 13 & $e2-hh2$ & 1.597 & 0.32 $\pm$ 0.01 \\
 & 8.8 & $e1-hh1$ & 1.555 & 0.20 $\pm$ 0.005 \\
 & 5.1 & $e1-hh1$ & 1.600 & 0.00 $\pm$ 0.04 \\
\hline
$\# 3$ (p337), x=0.28 & 10 & $e1-hh1$ & 1.544 & 0.26 $\pm$ 0.005 \\
\hline
$\# 4$ (11302), x=0.25 & 8.4 & $e1-hh1$ & 1.559 & 0.17 $\pm$ 0.01 \\
 & GaAs buffer & & 1.559 & 0.43 $\pm$ 0.005 \\
\hline
$\# 5$ (e294), x=0.32 & 8.8 & $e1-hh1$ & 1.555 & 0.21 $\pm$ 0.01 \\
 & GaAs buffer &  & 1.515 & 0.44 $\pm$ 0.001 \\
 & GaAs buffer &  & 1.529 & 0.44 $\pm$ 0.001 \\
 & GaAs buffer &  & 1.543 & 0.43 $\pm$ 0.006 \\
 & GaAs buffer &  & 1.553 & 0.42 $\pm$ 0.004 \\
\end{tabular}
\end{ruledtabular}
\end{table*}

The electron $g$-factor was evaluated from the frequency of the 
electron spin quantum beats corresponding to the Larmor 
precession frequency in magnetic field. A pump-probe technique 
with polarization sensitivity based on time-resolved Kerr 
rotation was used for detection of the spin beats (see e. g. 
Refs.~\onlinecite{Awsch, Crooker97}). A Ti:Sapphire laser emitting 1.8~ps pulses 
at a repetition rate of 75.6~MHz was tuned in resonance with the 
QW exciton transition. The pump beam was circularly polarized by 
means of an elasto-optical modulator operating at 50~kHz. The 
excitation density was kept as low as possible in the range from 
0.2 to 1~W/cm$^2$. The probe beam was linearly polarized, its 
intensity was about 20$\%$ of the pump beam. The rotation angle of 
the linearly polarized probe pulse reflected from the sample due 
to the Kerr rotation, was detected by a balanced diode detector 
and a lock-in amplifier. The time-resolved Kerr rotation signal 
as function of the delay between probe and pump pulses gives the 
evolution of the electron spin coherence generated by the pump.

\section{Experimental results}

A typical photoluminescence spectrum obtained for sample $\# 2$ containing four 
single QWs of different thickness is shown in Fig.~1. The emission from the 
GaAs buffer layer is presented by the dashed line. The luminescence from the 
QWs is dominated by recombination of excitons, whose energy increases for the 
narrow QWs due to confinement. The low energy shoulders of the excitonic lines 
are due to the negatively charged excitons (trions) recombination 
\cite{Buhmann95}. Trions consisting of two electrons and a hole are formed by a 
background electron and a photogenerated exciton.

\begin{figure}[hbt]
\includegraphics*[width=7cm]{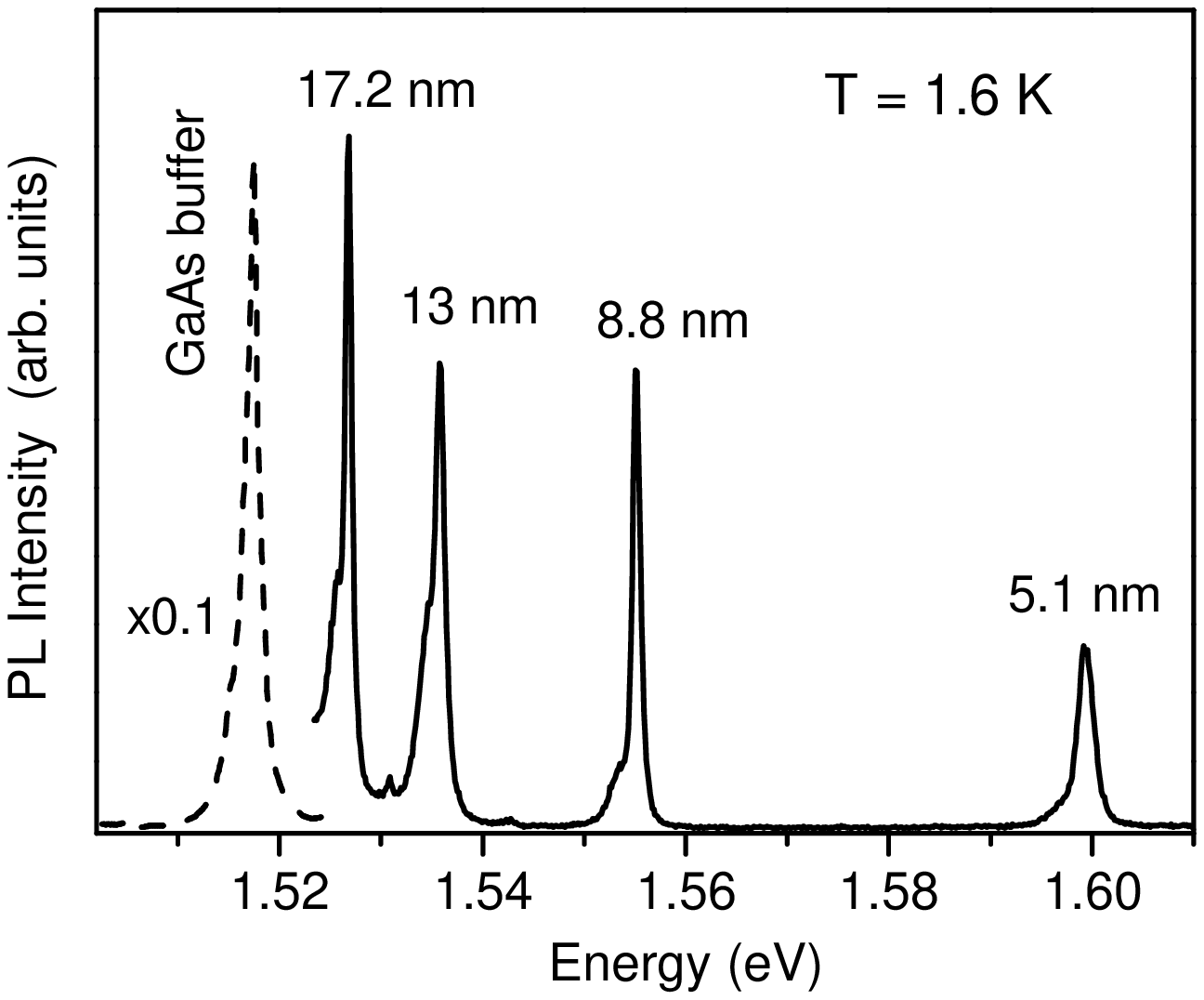}
\caption[] {Photoluminescence spectrum of a GaAs/Al$_{0.33}$Ga$_{0.67}$As 
structure containing four single QWs of different widths (sample $\# 2$). The 
emission from the GaAs buffer layer is shown by the dashed line. } \label{fig:1}
\end{figure}

\begin{figure}[hbt]
\includegraphics*[width=7cm]{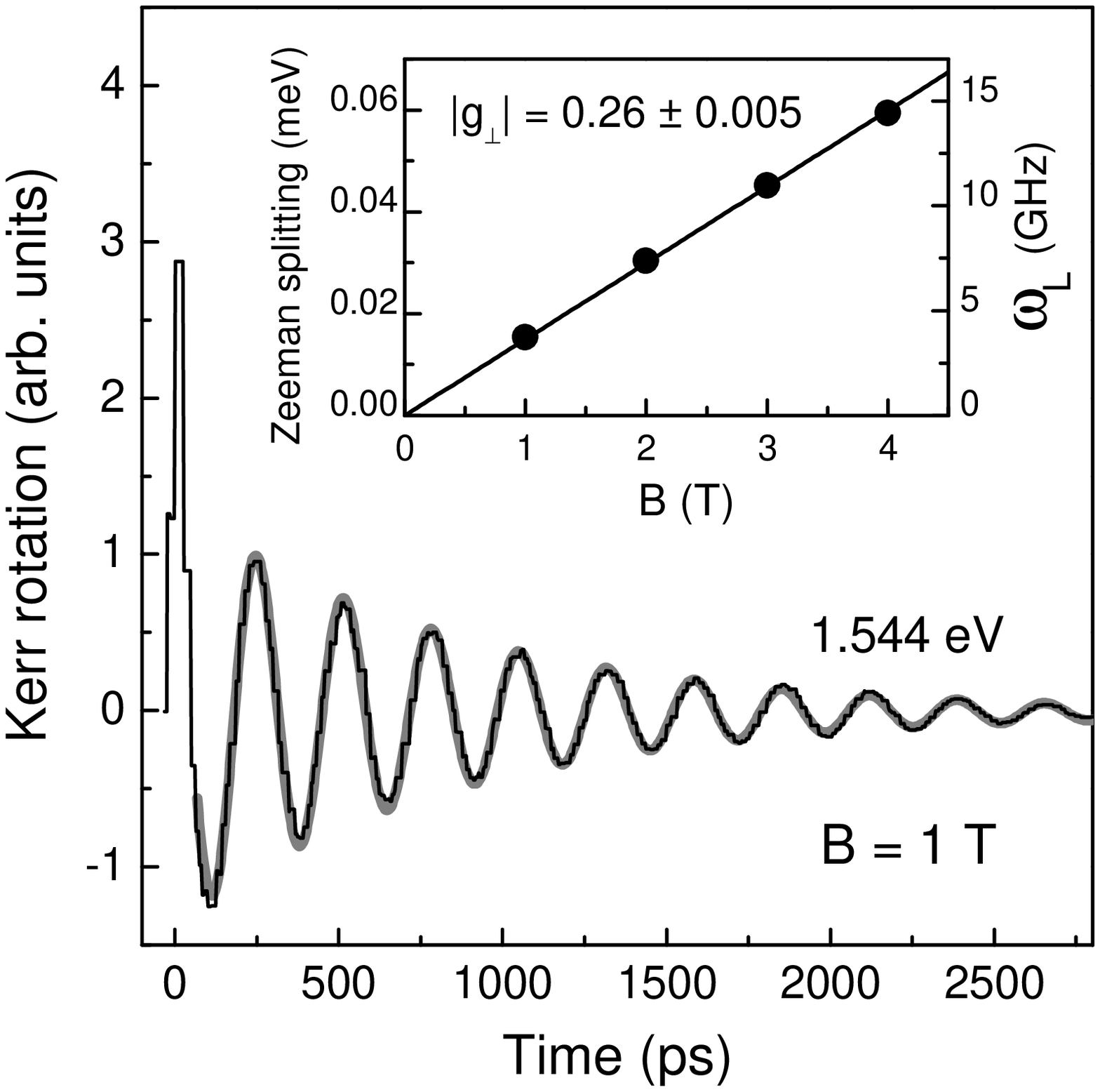}
\caption[] { Time-resolved Kerr rotation for a 10~nm QW (sample $\# 3$) in a 
magnetic field of 1~T. The black line gives the experimental data, the thick gray 
line is a fit after Eq.~(2) to the data using the parameters $\omega_L = 
3.76$~GHz  and $\tau = 880$~ps. The inset shows the Zeeman splitting $\Delta 
E$ (left scale) and the frequency of spin beats $\omega_L$ (right scale) as 
function of magnetic field. $T = 1.6$~K.} \label{fig:2}
\end{figure}

An example of time-resolved spin quantum beats in a 10~nm wide QW detected 
in a magnetic field of $B = 1$~T is shown in Fig.~2. The experimental data are 
plotted by the black line. The excitation energy of 1.544~eV is resonant with the 
exciton transition. The periodic oscillations of the Kerr signal are due to the 
precession of coherently excited electron spins about the magnetic field, with the 
oscillation period $T_L$ given by the spin splitting in the conduction band 
$\Delta E$. Therefore, the Larmor precession frequency $\omega_{L}=2 \pi 
/T_L$ allows precise determination of the transverse component of the electron 
$g$-factor $g_{\perp}$ by
\begin{equation}
\Delta E=\mu_B g_{\perp} B=\hbar \omega_L,
\label{eqn1}
\end{equation}
where $\mu_B$ is the Bohr magneton. Note that the sign of the $g$-factor cannot 
be determined, but only its absolute value. For that purpose, we fit the spin-beat 
dynamics by form for an exponentially damped oscillation. In case of a single 
frequency and a single decay time, which gives an appropriate description for 
most of the studied structures, this form is given by the following equation:
\begin{equation}
y(t)=A exp(-t/\tau) cos(\omega_L t),
\label{eqn2}
\end{equation}
with an amplitude $A$. $\tau$ is the decay time of spin coherence, which for an 
electron spin ensemble corresponds to the spin dephasing time $T_2^*$ 
\cite{Awsch, Zhu06}. An example for the result of a fitting to the experiment is 
given in Fig.~2 by the thick grey line. Here we exclude from the analysis the 
initial 15-30~ps after the pump pulse, which are contributed by the fast relaxation 
of holes \cite{Linder98, Amand97, Gerlovin04}.

In the inset, the determined Zeeman splitting is plotted as a function of magnetic 
field. The corresponding values of the Larmor frequency are also given on the 
right vertical axis. The slope of the B-linear dependence gives $|g_{\perp}| = 0.26 
\pm 0.005$. The decay time of the spin beats in Fig.~2 is 880~ps and is 
considerably longer than the radiative decay of resonantly excited excitons which 
does not exceed 100~ps in GaAs/Al$_x$Ga$_{1-x}$As QWs \cite{Deveaud}. 
Therefore, we conclude that the detected spin beats are provided by background 
electrons whose spin coherence is photogenerated by the trion formation 
\cite{Zhu06}. 

\begin{figure}[hbt]
\includegraphics*[width=7cm]{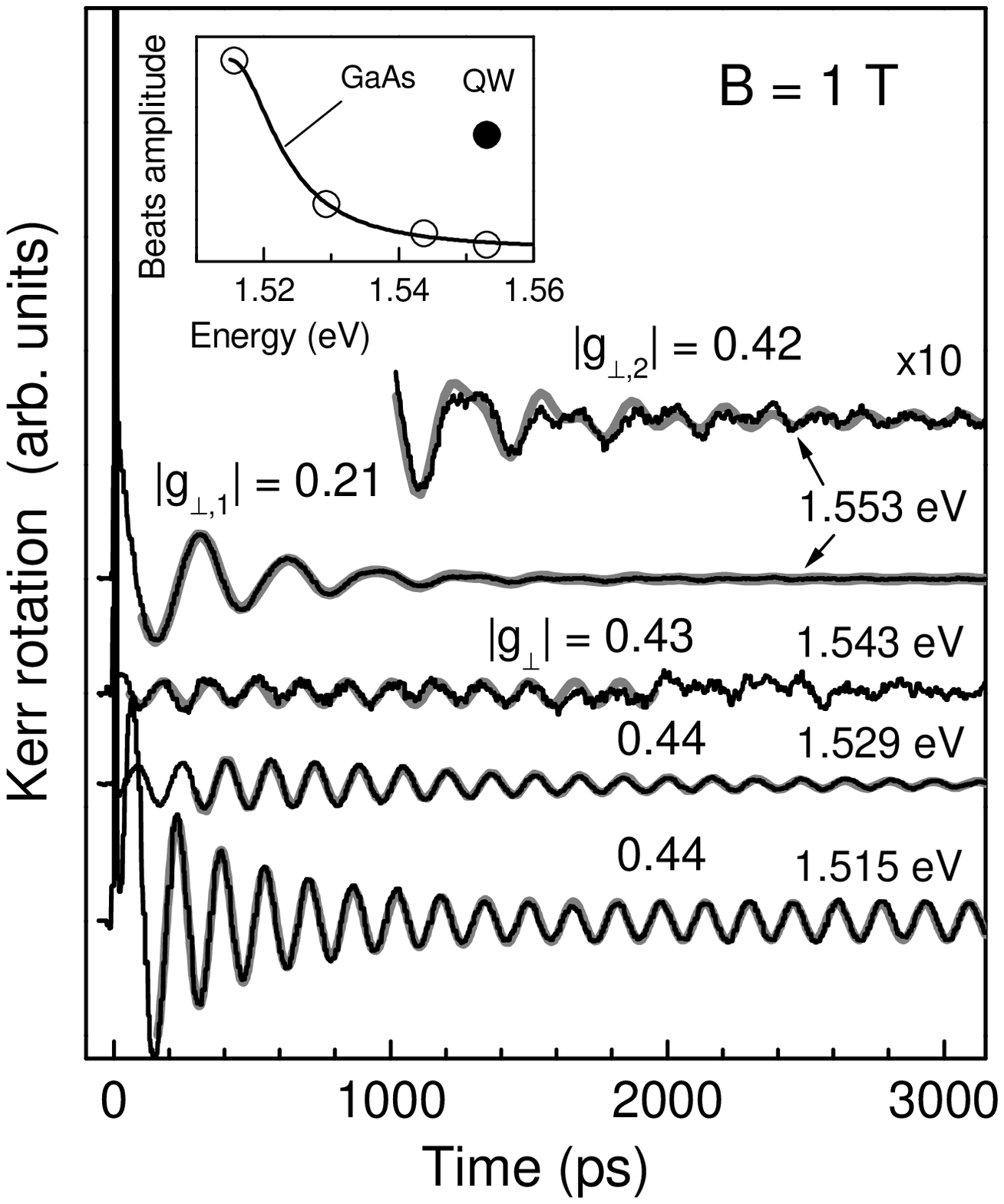}
\caption[] { Time-resolved Kerr rotation of an 8.8~nm QW (sample $\# 5$) 
measured at different excitation energies. The upper curve (1.533~eV) has been 
scaled to show clearly the beats from the GaAs buffer layer superimposed by the 
QW signal. The experimental data are shown by the narrow black lines and fit 
results by the forms discussed in the text are given by the thick gray lines. Inset: 
Relative intensities of the beats from the QW (closed circles) and the GaAs buffer 
layer (open circles). $T = 1.6$~K.} \label{fig:3}
\end{figure}	

Not in all cases the Kerr signals can be described by a single frequency and/or 
single decay time. Figure~3 illustrates a more complicated case observed in 
sample $\# 5$ for a 8.8~nm wide QW. The excitation energy of 1.515~eV is 
resonant with the exciton transition of the GaAs buffer layer and the observed 
oscillations with $|g_{\perp}|=0.44$ are in accord with the known $g$-factor 
value of bulk GaAs $g$(GaAs)$=-0.44$ \cite{Landolt}. The decay of the spin 
beats shows two characteristic times. The fast one of 360~ps may be associated 
with the exciton lifetime and the long one, which exceeds 3~ns describes the 
dephasing of the background electrons in the buffer layer \cite{Kikkawa98}. The 
resonant excitation into the exciton states of GaAs results in the largest amplitude 
of the Kerr signal. However, pronounced spin beats can be observed when the 
excitation energy is detuned from the resonance condition and shifted further up 
by 27~meV to the energy of 1.543~eV, which is still below any QW resonance. 
The Kerr signal at these energies is provided by the exciton-polariton reflection 
spectrum \cite{Tredicucci}, which is governed by the quantization of polaritons in 
the GaAs buffer layer. The spin beats period is independent of the excitation 
energy, as it is determined by spin oriented carriers, which relax from the excited 
states to the bottom of the conduction band and precess there with 
$|g_{\perp}|=0.44$. The amplitude of the Kerr signal, however, decreases, as 
shown by the open circles in the inset.

An increase of the laser energy to 1.553~eV brings it in resonance with the QW 
exciton. The beats period increases about twice corresponding to $|g_{\perp , 
1}|=0.21$. The spin beats are damped with a decay time $\tau =410$~ps. At 
longer delays exceeding 1000~ps the oscillation picture becomes irregular, 
suggesting that at least two periodic processes overlap. The signal was fitted by an 
equation accounting for two frequencies with different decay times: 
\begin{eqnarray}
y(t)=A_1 exp(-t/\tau_{1}) cos(\omega_{L,1} t + \varphi_1) \nonumber \\+ 
A_2 exp(-t/\tau_{2}) cos(\omega_{L,2} t + \varphi_2).
\label{eqn3}
\end{eqnarray}

As one can see from the fit in Fig.~3 shown by the grey lines, the experiment at 
longer delays can be described by spin beats of QW electrons with $|g_{\perp , 
1}|=0.21$ superimposed with $|g_{\perp , 1}|=0.42$ beats. The latter beats can be 
attributed to electrons precessing in the GaAs buffer. The relative intensities of 
the amplitudes $A_1$ and $A_2$ are given in the inset of Fig.~3.

To obtain further insight, a 13~nm wide QW in sample $\# 2$ has been excited 
resonantly with three optical transitions corresponding to the ground state of the 
heavy-hole exciton ($e1-hh1$), of the light-hole exciton ($e1-lh1$) and of the 
exciton related to the second confined electron and levels ($e2-hh2$). For all 
cases spin beats with almost the same period corresponding to $|g_{\perp}|=0.33$  
have been found (see Table~\ref{tab:table1}). As the value of the electron $g$-
factor should strongly depend on the electron energy, we therefore conclude that 
the Kerr signal is provided by electrons which relax to the bottom of the 
conduction band shortly after the photogeneration and precess there. It is 
remarkable that the $e1$ electron precession can be accessed through the $e2-
hh2$ optical transition.

\begin{figure}[hbt]
\includegraphics*[width=7cm]{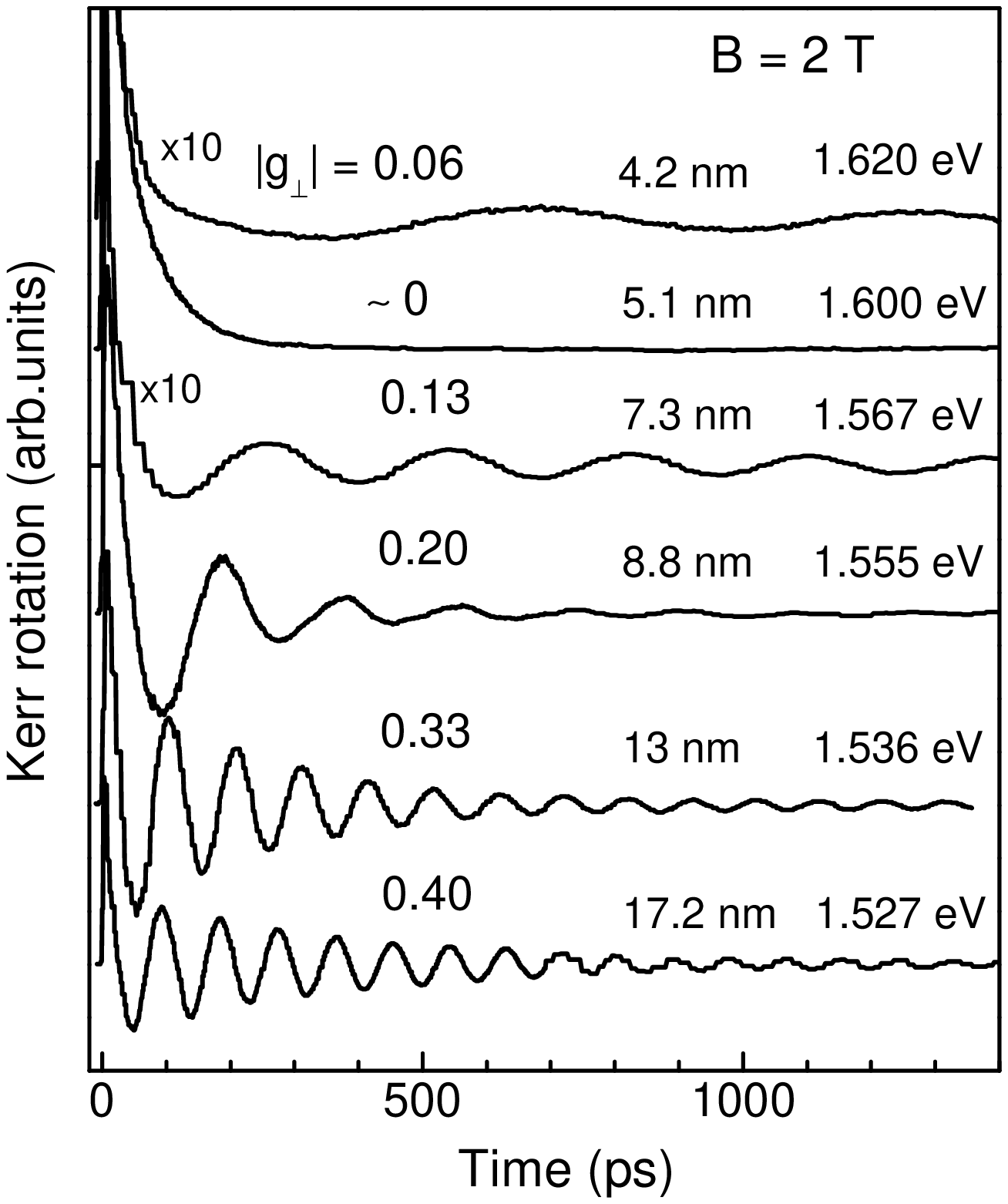}
\caption[] { Time-resolved Kerr rotation in QWs of different thicknesses (samples 
$\# 1$ and $\# 2$) measured in a magnetic field of 2~T. Well width, excitation 
energy and $g$-factor value are indicated for each signal. $T = 1.6$~K.} 
\label{fig:4}
\end{figure}

After having given some insight into the general features of the experimental data 
and their analysis, we turn to the problem of the $g$-factor dependence on the 
carrier confinement conditions. Time-resolved Kerr rotation signals detected for 
the laser energy resonant with the exciton optical transitions in QWs of different 
width are given in Fig.~4. One can see that the spin beats frequency, which is 
proportional to $|g_{\perp}|$ of the conduction band electrons, decreases with 
decreasing well width. Oscillations cannot be resolved in a 5.1~nm QW with the 
exciton energy at 1.600~eV. A further decrease of the QW width down to 4.2~nm 
restores the spin beat oscillations. The determined $|g_{\perp}|$ values are given 
in the figure and are collected also in Table~\ref{tab:table1}.

\begin{figure}[hbt]
\includegraphics*[width=7cm]{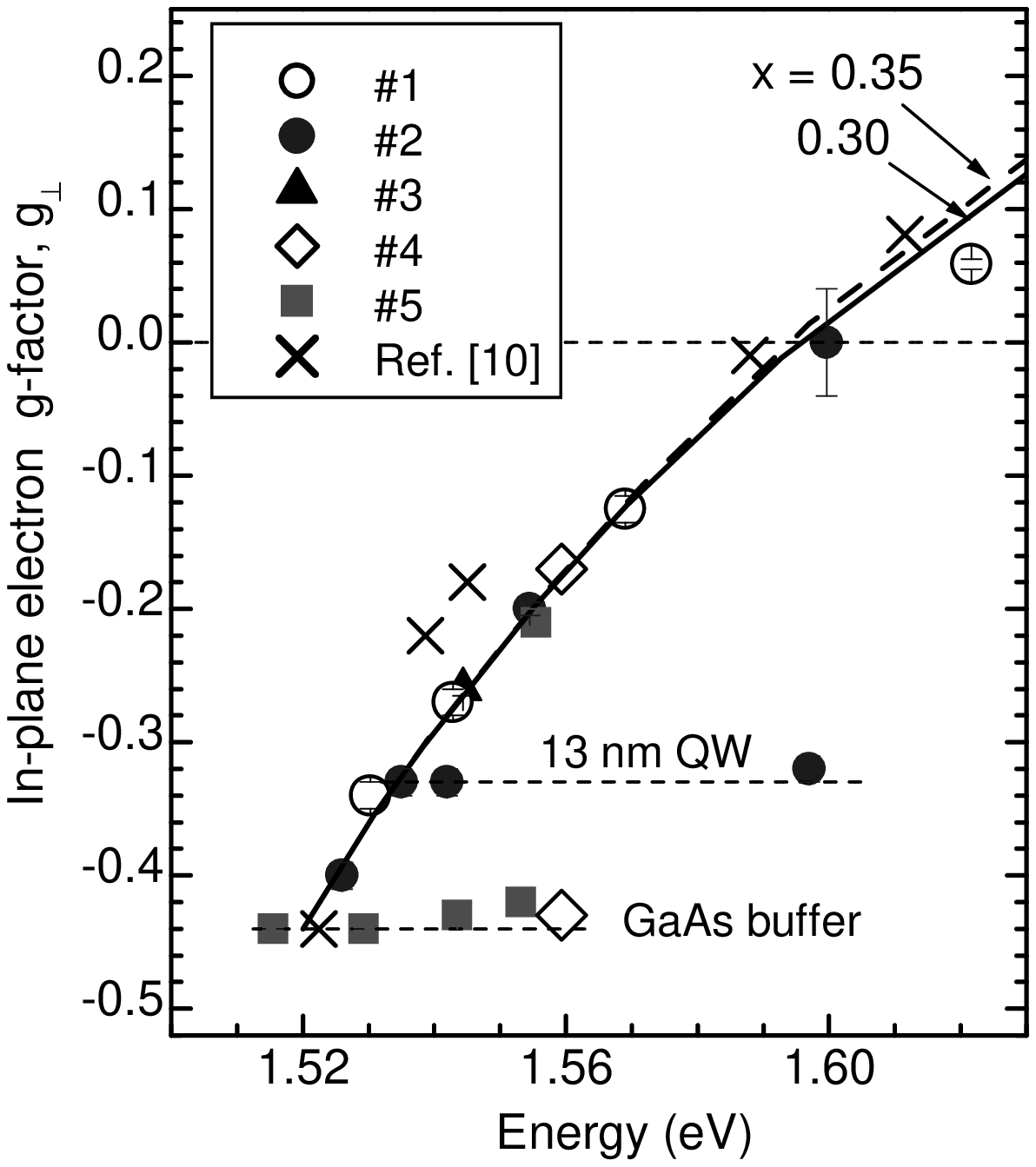}
\caption[] { Dependence of the transverse component of the electron $g$-factor 
on excitation energy. Symbols are experiment, solid and dashed lines are 
calculations for Al contents of $x = 0.3$ and 0.35, respectively. Horizontal dashed 
lines are guides to the eye. $T = 1.6$~K. } \label{fig:5}
\end{figure}

The experimental values for $|g_{\perp}|$ are plotted in Fig.~5 as function of 
excitation energy. Most of the data, except for some results for a 13~nm QW and 
the GaAs buffer which were measured for non-resonant excitation, were detected 
when the laser was resonant with the exciton state $e1-hh1$. Measurements on 
different structures confirm the monotonic decrease of the $g$-factor absolute 
value with increasing energy leading to a sign reversal at 1.600~eV. Our results 
are in good agreement with the data of Snelling et al. \cite{Snelling91} obtained 
by the optical orientation technique. They are also in good agreement with model 
calculations for Al contents of $x = 0.3$ and 0.35 shown by the solid and the 
dashed lines, respectively. Details of these calculations are given below in 
Sec.~\ref{Sec:calc}. We note that the calculated dependencies are plotted as 
function of energy of the optical transition between the confined carrier levels 
$e1$ and $hh1$ without accounting for the exciton binding energy. The latter 
may shift these curves to lower energy by about 8~meV in wide QWs of about 
20~nm and by about 13~meV in 4~nm QWs \cite{Bimberg}.

\section{ Calculation of the electronic $g$-factor and comparison 
with experiment.} \label{Sec:calc}

Here we present results of model calculations for the longitudinal and transverse 
components of the electronic $g$-factor in GaAs/Al$_x$Ga$_{1-x}$As QWs for 
a wide range of well widths from 1 to 30~nm and Al contents $0<x<0.45$, 
which control the barrier height. In Fig.~6 the calculated $g$-factors are shown as 
function of the e1-hh1 optical transition energy. This provides a convenient 
comparison with experiment, as the $g$-factor is addressed in dependence of an 
easily measurable quantity. 

\begin{figure}[hbt]
\includegraphics*[width=7cm]{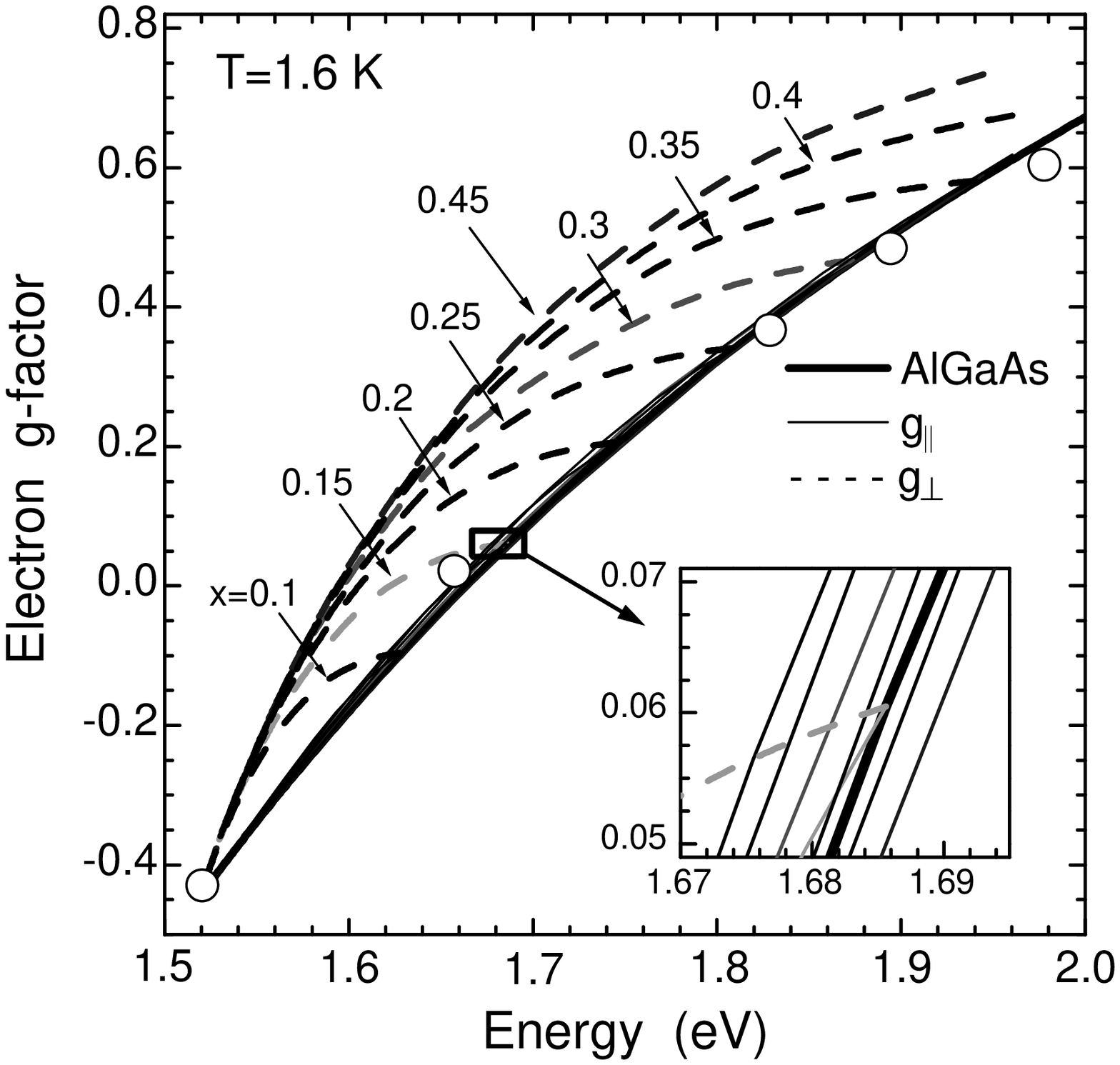}
\caption[] { Longitudinal (solid lines) and transverse (dashed lines) components 
of the electron $g$-factor as function of the energy of the optical transition in 
GaAs/Al$_x$Ga$_{1-x}$As QWs, calculated for various barrier compositions 
and different QW widths. Open circles show experimental data for 
Al$_x$Ga$_{1-x}$As alloys taken from Ref.~\onlinecite{Hermann}. Inset details 
dependences for the longitudinal $g$-factor, which closely follow the dependence 
for Al$_x$Ga$_{1-x}$As.} 
\label{fig:6}
\end{figure}

	To prepare Fig.~6, it was necessary, first, to calculate accurately the 
lowest quantized states of electrons and heavy holes for each set of 
heterostructure parameters, and, second, to evaluate the $g$-factor tensor for the 
calculated electron state. To achieve advanced accuracy in the energy levels, we 
applied the Kane multiband Hamiltonian, accounting exactly for the coupling 
between the lowest conduction band $\Gamma_6$ and the upper valence bands 
$\Gamma_8$ and $\Gamma_7$, and retaining also all remote band terms that 
notably affect the dispersion of the relevant quasi-particles in conduction and 
valence band. Details of the computational procedure were presented elsewhere 
\cite{Kiselev_PRB01,Kiselev_APL02}. Then by directly following the 
prescriptions in Ref. \onlinecite{Kiselev_PhysStatSol99}, two independent components 
of the $g$-factor tensor at the bottom of the first electron subband in the QW 
structure can be explicitly calculated: the in-plane $g$-factor $g_{\perp}$  using 
Eq.~(6) in Ref.~\onlinecite{Kiselev_PhysStatSol99} and the $g$-factor along the 
growth direction $g_{| |}$ with help of Eq.~(10) in 
Ref.~\onlinecite{Kiselev_PhysStatSol99}.

\begin{table}
\caption{\label{tab:table2} Band structure parameters for GaAs and 
Al$_{0.35}$Ga$_{0.65}$As. Also included is the type of interpolation procedure 
for the Al$_x$Ga$_{1-x}$As alloys. The data are taken from 
Ref.~\onlinecite{Landolt}.}
\begin{ruledtabular}
\begin{tabular}{c|ccc}
& GaAs &  Al$_{0.35}$Ga$_{0.65}$As & Interpolation \\
\hline
$\Delta_{SO}$~(eV) & 0.341 & 0.32 & basic, linear \\
$2P_{cv}^2/m_0$~(eV) & 28.9 & 26.7 & basic, linear for $P_{cv}$ \\
$m_{bulk}$ & 0.067 $m_0$ & 0.09 $m_0$ & composite \\
$g_{bulk}$ & -0.44 & 0.58 & composite \\
$m_{hh}$ & 0.45 $m_0$ & 0.45 $m_0$ & constant \\
\end{tabular}
\end{ruledtabular}
\end{table}

The parameters characterizing the band structure of the Al$_x$Ga$_{1-x}$As 
alloy are known to a great definiteness. Numeric values for all applicable 
quantities are collected in Table~\ref{tab:table2} for GaAs and 
Al$_{0.35}$Ga$_{0.65}$As (the data are taken from Ref. \onlinecite{Landolt}). The 
following scheme for parameter evaluation has been adopted for an arbitrary Al 
composition $x$: for most of the parameters (named {\em basic} in 
Table~\ref{tab:table2}), we use a simple linear interpolation (extrapolation 
for $x > 0.35$). This group contains the interband momentum matrix element 
$P_{cv}$ and the spin-orbit splitting $\Delta_{SO}$ in the valence band. 
Rigorously speaking, higher accuracy can probably be achieved if also the bowing 
constants for the interpolation curves were known, but for the studied range $0 < 
x < 0.45$ linear interpolations should be sufficient. As the only notable exception, 
we use the common interpolation equation with bowing term for the 
Al$_x$Ga$_{1-x}$As band gap \cite{Landolt}:
\begin{equation}
E_g(x)=1.519+1.04x+0.47x^2[eV].
\label{eqn4}
\end{equation}

Some parameters are obviously model derivatives and we denote them {\em 
composite}: the bulk electron effective mass $m_{bulk}$ and the bulk $g$-factor 
$g_{bulk}$ are defined by a subset of basic quantities. These composite 
parameters are not expected to follow a linear interpolation law, so we apply a 
different approach: when we need the value of a composite parameter for some 
alloy composition, we directly {\em calculate} it from {\em interpolated} basic 
parameters. For example \cite{Roth},
\begin{equation}
g_{bulk}=g_0-
\frac{4}{3}\frac{P_{cv}^2}{m_0}\frac{\Delta_{SO}}{E_g(E_g+\Delta_{SO})}
+\delta g_{remote}
\label{eqn5}
\end{equation}
where $m_0$ is mass of free electron in vacuum and $\delta g_{remote}$ is a 
correction due to the higher lying bands. First, we use the specific numerical 
values of $g_{bulk}$ from Table~\ref{tab:table2} and Eq.~(5) to evaluate 
$\delta g_{remote}$ (which is treated here as {\em basic} and interpolated 
when necessary). Then, we apply Eq.~(5) again to calculate $g_{bulk}$ for 
arbitrary alloy composition from a complete set of interpolated basic parameters. 
The heavy hole 
effective mass $m_{hh}$ is taken to be independent of the alloy composition. A 
ratio $\Delta E_C/\Delta E_V =60/40$ for the offsets in the conduction and 
valence bands was taken for the calculations \cite{Landolt_heterostructures}.
	
Let us now describe in detail the results summarized in Fig.~6. Here the 
two independent components of the $g$-factor tensor are shown for the electron 
confined in QWs with different barrier composition $x$ as function of the $e1-
hh1$ optical transition energy. The components for magnetic field applied parallel 
($g_{| |}$) and perpendicular ($g_{\perp}$) to the structure growth axis are given 
by the solid and dashed lines, respectively. For each composition of the barrier 
material $g_{\perp} > g_{| |}$. The lowest energy for the optical transition is 
achieved for an infinitely wide QW, for which the energy asymptotically 
approaches that of bulk GaAs (1.519~eV at $T = 1.6$~K) and both $g$ tensor 
components converge to meet the electron $g$-factor value in the bulk: 
$g$(GaAs)$=-0.44$. Therefore, for each $x$ value the curves for the two $g$-
factor components form a petal with the {\em root} corresponding to the principal 
band gap and the $g$-factor of bulk GaAs and the {\em tip} corresponding to the 
values of the Al$_x$Ga$_{1-x}$As barriers (limit of a very narrow QW). As the 
band gap and the $g$-factor both grow monotonously with the Al composition, 
the petal size also increases and its tip draws the line $g_{bulk}(E_g)$ 
corresponding to the bulk $g$-factor value for a range of alloy compositions 
(shown in Fig.~6 by a thick solid line, open circles are experimental data for 
Al$_x$Ga$_{1-x}$As alloys \cite{Hermann}).

Implicitly, the thick solid line in Fig.~6 is defined by Eq.~(5): assuming linear 
interpolations for $P_{cv}$, $\Delta_{SO}$ and $\delta g_{remote}$, and the 
composition dependence for $E_g(x)$ given by Eq.~(4), $g_{bulk}(E_g)$ can be 
easily recovered analytically as a series expansion. For the GaAs/Al$_x$Ga$_{1-
x}$As heterosystem, the first terms in the expansion are \cite{Landolt}:
\begin{equation}
g_{bulk}(E_g) \approx  -0.445+3.38(E_g-1.519)-2.21(E_g-1.519)^2,
\label{eqn6}
\end{equation}
with $E_g$ measured in eV.
	
Quite remarkably, the growth direction component $g_{| |}(E_{e1-hh1})$ follows 
{\em very closely} the Al$_x$Ga$_{1-x}$As dependence $g_{bulk}(E_g)$ for an 
arbitrary barrier composition. The inset in Fig.~6 shows a close-up of the energy 
dependence of $g_{| |}$ for different $x$. It illustrates that the $g_{| |}$ values 
for the whole range of $x$ from 0 up to 0.45 basically coincide, besides minor 
deviations, with the dependence for bulk Al$_x$Ga$_{1-x}$As. 
	
Although $g_{\perp}$ follows generally the same trend, it deviates 
notably from the alloy dependence. This deviation is caused by the structural 
anisotropy in the structure: therefore it depends on the barrier height and strength 
of spatial confinement. However, for the studied GaAs/Al$_x$Ga$_{1-x}$As 
QWs the maximum deviation of $g_{\perp}$ from the bulk dependence is always 
moderate, and never exceeds 0.3. Also, in the immediate vicinity of the petal root 
(corresponding to the case of very weak spatial confinement and negligible 
penetration of the electron state into barriers) some universality with respect to 
different barrier compositions can be spotted for $g_{\perp}$.

\section{ Discussion and conclusions}
In order to assess the meaning and the validity of the universal dependence of the 
$g$-factor on the heterostructure band gap, let us first consider a hypothetical 
alloy heterosystem A$_x$B$_{1-x}$. We assume that the system can be 
accurately described by the Kane model, in which the two {\em basic} parameters 
(i) the interband matrix element $P_{cv}$ and (ii) the spin-orbit splitting of 
valence band $\Delta_{SO}$ are plain independent of the composition 
(consequently, these quantities will be equal in the well and barrier layers). 
Another assumption is that (iii) the valence band offset is exactly zero. Analysis 
shows, that for such a remarkable heterosystem, a {\em truly} universal dependence 
of the electron $g$-factor on the $e1-hh1$ energy in the QW structure is expected.  
Moreover, no 
$g$-factor anisotropy would be expected. We would like to note that similar 
universality should also reappear when the barriers in the heterostructures are 
extremely high, preventing considerable penetration of the electron wave function 
into the barriers. 
	
In reality, conditions (i)-(iii) are not satisfied, so that the degree of the 
parameter mismatch (along with the extent of the penetration of a confined 
electron into the barrier layers) governs the deviations from the universal 
behaviour when one changes barrier composition and QW width. The matrix 
element $P_{cv}$ is modified only moderately from one semiconductor to 
another and the modulation in the value of $\Delta_{SO}$ is almost negligible in 
cation-substituted materials, including the Al$_x$Ga$_{1-x}$As alloys. 
However, it can be considerable in anion-substituted solid solutions \cite{Landolt}. 
Condition 
(iii) appears to be most vulnerable, as the valence band offset accounts for about 
40$\%$ of the difference in band gaps for the well and barrier materials in 
GaAs/Al$_x$Ga$_{1-x}$As heterostructures, but some approximate universality 
can be reasonably expected. The profound universality in the behavior of the 
growth direction component of the electron $g$-factor in GaAs/Al$_x$Ga$_{1-
x}$As QWs, even though it is not exact, obviously exceeds these expectations. A 
detailed numerical analysis shows that it appears due to a fragile and delicate 
balance in the dependence of the material parameters on composition: when the 
band gap increases with Al content, slight reductions of $P_{cv}$ and 
$\Delta_{SO}$ {\em counter play} and {\em mostly compensate} the effect of 
moderate but nonzero valence band offset in the type~I band alignment at the 
heterointerface, the particular experimental value of $m_{hh}$ also helps.
	
To conclude the discussion, approximate universalities in the $g$-factor 
behavior can be expected in general for cation-substituted alloy heterostructures 
with a type I band alignment and small-to-moderate valence band offsets. 
Heterosystems with these properties are known to include a number of important 
III-V and II-VI ternary semiconductors. However, one should be cautious about 
an indiscriminate extension of these conclusions to arbitrary materials. 

In summary, the electronic $g$-factor has been studied experimentally and 
theoretically in GaAs/Al$_x$Ga$_{1-x}$As QWs for a wide range of well widths 
and Al contents. The results are represented as a $g$-factor dependence on the 
energy of the $e1-hh1$ optical transition in QWs. A remarkable universality has 
been established for the $g$ tensor components along the structure growth axis 
$g_{| |}$. The deviation of the transverse components $g_{\perp}$ from this 
universal behavior is controlled by the structure anisotropy. 

\section*{ACKNOWLEDGMENTS}
We appreciate fruitful discussions with E. L. Ivchenko and with I. V. Ignatiev. 
We are thankful to R. T. 
Harley for providing us additional information about his samples. This work has 
been supported by the BMBF program "nanoquit", by the ISTC grant 2679 and by 
the RFBR grant 06-02-17157. Research 
stays of IAY in Dortmund have been financed by the 
Deutsche Forschungsgemeinschaft via Graduiertenkolleg 726 "Materials and 
Concepts for Quantum Information Processing" and grants Nos. 436 RUS 
17/98/05 and 436 RUS 17/69/06. AAK acknowledges partial financial support 
from DARPA and ONR.

\end{document}